\documentclass[reprint,two-column,double-spaced,groupedaddress,superscriptaddress,pra]{revtex4-1}

\usepackage{mathrsfs}
\usepackage{enumerate}
\usepackage{graphicx}
\usepackage{physics}
\usepackage[colorlinks]{hyperref}

\begin{document}


\title{Identification of networking quantum teleportation \\ on 14-qubit IBM universal quantum computer}


\author{Ni-Ni Huang}
\affiliation{Department of Engineering Science, National Cheng Kung University, Tainan 701, Taiwan}
\affiliation{Center for Quantum Frontiers of Research $\&$ Technology, NCKU, Tainan 701, Taiwan}
\author{Wei-Hao Huang}
\affiliation{Department of Engineering Science, National Cheng Kung University, Tainan 701, Taiwan}
\affiliation{Center for Quantum Frontiers of Research $\&$ Technology, NCKU, Tainan 701, Taiwan}
\author{Che-Ming Li}
\email{cmli@mail.ncku.edu.tw}
\affiliation{Department of Engineering Science, National Cheng Kung University, Tainan 701, Taiwan}
\affiliation{Center for Quantum Frontiers of Research $\&$ Technology, NCKU, Tainan 701, Taiwan}
\affiliation{Center for Quantum Technology, Hsinchu 30013, Taiwan}


\date{\today}

\begin{abstract}
Quantum teleportation enables networking participants to move an unknown quantum state between the nodes of a quantum network, and hence constitutes an essential element in constructing large-sale quantum processors with a quantum modular architecture. Herein, we propose two protocols for teleporting qubits through an $N$-node quantum network in a highly-entangled box-cluster state or chain-type cluster state. The proposed protocols are systematically scalable to an arbitrary finite number $N$ and applicable to arbitrary size of modules. The protocol based on a box-cluster state is implemented on a 14-qubit IBM quantum computer for $N$ up to 12. To identify faithful networking teleportation, namely that the elements on real devices required for the networking teleportation process are all qualified for achieving teleportation task, we quantify quantum-mechanical processes using a generic classical-process model through which any classical strategies of mimicry of teleportation can be ruled out. From the viewpoint of achieving a genuinely quantum-mechanical process, the present work provides a novel toolbox consisting of the networking teleportation protocols and the criteria for identifying faithful teleportation for universal quantum computers with modular architectures and facilitates further improvements in the reliability of quantum-information processing.
\end{abstract}


\maketitle

\section{INTRODUCTION}

\begin{figure*}[t]
\includegraphics[width=17cm]{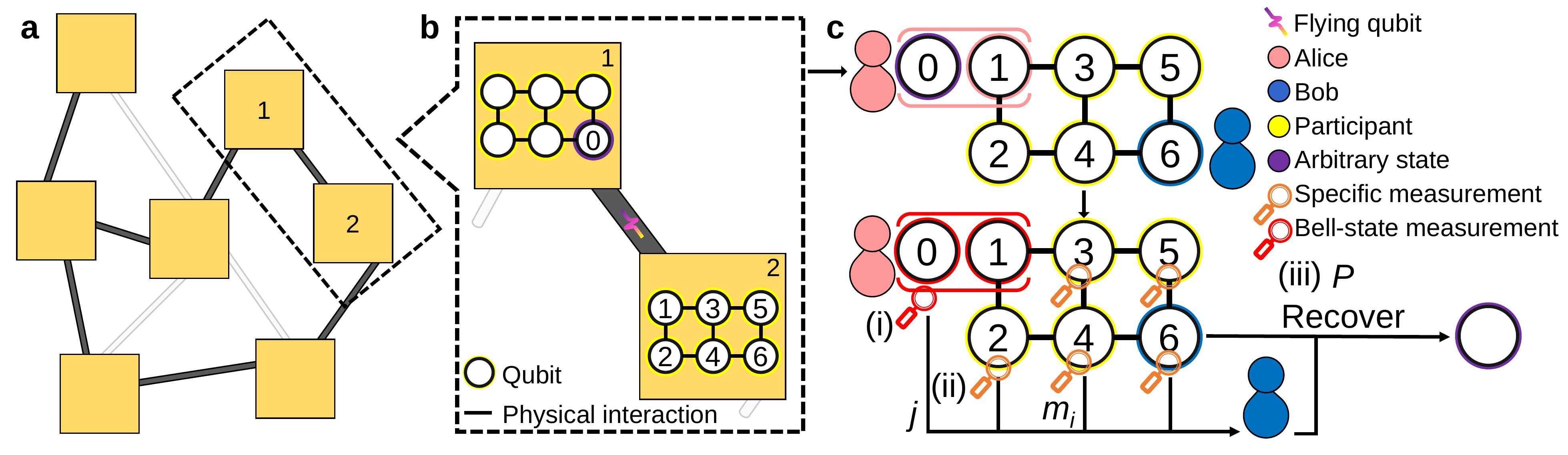}
\caption{\label{fig:fig_1} Networking teleportation in a modular architecture. (\textbf{a}) Modular architecture of the quantum network. Each module is regarded as a node of the quantum network and consists of qubits \cite{monroe2013scaling, childress2014atom, monroe2014large, hucul2015modular, chou2018deterministic, leung2019deterministic}. The modules transmit quantum states to one another by performing intra- and inter-module operations \cite{monroe2013scaling, devoret2013superconducting, monroe2014large, hucul2015modular, narla2016robust, brecht2016multilayer, chou2018deterministic, leung2019deterministic}. The architecture can be extended to realize a universal quantum computer by teleporting a controlled-$Z$ (CZ) gate \cite{chou2018deterministic}. (\textbf{b}) Transmission of quantum information among different modules. Due to the isolation between modules, quantum teleportation \cite{monroe2013scaling, monroe2014large, hucul2015modular, narla2016robust, brecht2016multilayer, chou2018deterministic, leung2019deterministic} is utilized to transport an arbitrary quantum state of qubit 0 (purple) from module 1 to module 2 consisting of qubits 1--6. In particular, qubit 0 is entangled with a flying photon qubit \cite{devoret2013superconducting, childress2014atom, narla2016robust, chou2018deterministic} (gradient purple) via a CNOT-like operation on Alice's side \cite{narla2016robust}. Alice then measures qubit 0 on a specific basis to enable the flying qubit to carry the information regarding the quantum state of qubit 0 to be transported via remote state preparation \cite{bennett2001remote}. (\textbf{c}) Networking teleportation protocol based on a 6-qubit box-cluster state $\ket{C_{b, 6}}$. The protocol aims to teleport the arbitrary state of qubit 0 from Alice to Bob using a shared entangled state $\ket{C_{b, 6}}$ among Alice, the participants and Bob. The main steps of the protocol are as follows: (\romannumeral1) Bell-state measurement (red) on Alice's flying photon qubit which carries the information of the quantum state of qubit 0 and qubit 1; (\romannumeral2) specific local measurements (orange) on participants' qubits 2--5; (\romannumeral3) appropriate unitary transformations $P$ on Bob's qubit $N$ to recover the initial input state (purple) in accordance with the measurement results informed by Alice and the participants. (See text for detailed procedures.)} 
\end{figure*}

Quantum teleportation provides a method for transporting unknown quantum states between remote systems based on shared entanglement and quantum measurements~\cite{bennett1993teleporting}. Teleportation constitutes the fundamental element required to perform a wide range of quantum computation and quantum information tasks in a quantum network \cite{kimble2008quantum, ritter2012elementary, hucul2015modular, wehner2018quantum, pirker2018modular, chou2018deterministic, jing2019entanglement, yamasaki2019distributed}. In particular, to construct large-scale quantum computing processors with a modular architecture \cite{monroe2013scaling, devoret2013superconducting, childress2014atom, monroe2014large, hucul2015modular, narla2016robust, brecht2016multilayer, chou2018deterministic, leung2019deterministic} (Fig.~\ref{fig:fig_1}a), ideal teleportation is required to connect the various modules within the network \cite{monroe2013scaling, monroe2014large, hucul2015modular, narla2016robust, brecht2016multilayer, chou2018deterministic, leung2019deterministic} (Figs.~\ref{fig:fig_1}b, c). Notably, such an ideal quantum process is also essential for the modularization of quantum networks in which spatially-separated quantum nodes communicate across different modules \cite{pirker2018modular}. 
 
Recently, IBM launched the IBM Q Experience, which makes universal quantum computers accessible to the general public through cloud service \cite{ibmq}. IBM has also developed Qiskit \cite{Qiskit} to provide users with the tools required to run their quantum programs on prototype quantum devices and simulators. IBM Q Experience provides an online platform for the experimental testing of the fundamentals of quantum physics \cite{alsina2016experimental, wang201816, mooney2019entanglement, morris2019non, ku2019experimental} and a wide variety of applications in quantum information theory \cite{devitt2016performing, fedortchenko2016quantum, sisodia2017design, behera2017experimental, behera2018designing}. However, while IBM has built both 20-qubit and 50-qubit quantum processors \cite{ibmq}, a comprehensive characterization of the networking teleportation process for future modular use on IBM quantum devices is still lacking.

Accordingly, we present herein a toolbox for examining the performance of IBM quantum computers, where the networking teleportation protocol is executed. We firstly propose two systematically extensible networking teleportation protocols for a network consisting of $N$ parties based on either a box-cluster state or a chain-type cluster state \cite{briegel2001persistent, raussendorf2001one}. The proposed protocols possess applicability to arbitrary finite size of modules and the adaptability to the benchmark provided by a generic classical process model \cite{hsieh2017quantifying}. We then implement the proposed protocol based on the $N$-qubit box-cluster state on a 14-qubit quantum processor named \textit{ibmq\_16\_melbourne} for $N$ up to 12. 

A generic classical-process model providing the strictest criteria in order to rule out any classical strategies of mimicry of teleportation \cite{hsieh2017quantifying} is utilized for assessing the performance of the real quantum device, on which the proposed networking teleportation protocol is conducted. Through the generic classical-process model, one can identify whether the experimental networking teleportation process is faithful. In the case that the experimental process is identified as faithful, all the elements on real quantum devices required in the networking teleportation process are identified as all qualified for use. The model is defined as input states satisfying the assumption of realism and their evolutions to output states that can be reconstructed as a density operator, where this evolution conforms to classical stochastic theory. Existing identification methods utilize the state characteristics to verify the teleportation of IBM Q \cite{devitt2016performing, fedortchenko2016quantum, sisodia2017design}, solid-state systems \cite{gao2013quantum, bussieres2014quantum, pfaff2014unconditional}, trapped atoms \cite{riebe2004deterministic, nolleke2013efficient}, photonic qubits \cite{ursin2004communications, jin2010experimental, yin2012quantum, metcalf2014quantum, wang2015quantum}, atomic ensembles \cite{chen2008memory, krauter2013deterministic} and satellite-based systems \cite{ren2017ground, yin2017satellite}. By contrast, the present study provides a novel toolbox consisting of the two scalable $N$-qubit networking teleportation protocols and the criteria for identifying faithful teleportation.

\section{Cluster states}To implement the proposed networking teleportation protocol, it is first necessary to generate highly-entangled multipartite states (so-called cluster states) \cite{briegel2001persistent, raussendorf2001one}. Cluster states with multipartite quantum correlations are considered to be the significant source: basic building block when constructing general modular architectures for quantum networks \cite{pirker2018modular}. 

An $N$-qubit cluster state $\ket{C}$ can be generated by applying the controlled-$Z$ (CZ) gates with a specified configuration to the initial states, i.e.,
\begin{eqnarray}
\label{eq:1}
\ket{+}^{\otimes N}={H}^{\otimes N}\ket{0}^{\otimes N},
\end{eqnarray}
where $\ket{+}=(\ket{0}+\ket{1})/\sqrt{2}$ and $H=(\ket{0}\bra{0}+\ket{0}\bra{1}+\ket{1}\bra{0}-\ket{1}\bra{1})/{\sqrt{2}}$ is the Hadamard transformation ($H$). The state vector $\ket{C}$ can then be written in the form
\begin{eqnarray}
\label{eq:2}
\ket{C}=\prod_{(a') {\in} \mathcal{I}(a)}{\rm{CZ}}_{({a,a'})}\ket{+}^{\otimes N},
\end{eqnarray} 
where $\mathcal{I}$($a$) is the set of qubits that physically interact with qubit $a$, and ${\rm{CZ}}_{a,a'} = \ket{0}\bra{0}_a \otimes {I}_{a'} + \ket{1}\bra{1}_a \otimes {Z}_{a'}$ denotes the ${\rm{CZ}}$ gate acting on the control qubit $a$ and target qubit $a'$. Here, $I$ is a 2-dimensional identity matrix, and ${Z = (\ket{0}\bra{0}-\ket{1}\bra{1})}$ and ${X = (\ket{+}\bra{+}-\ket{-}\bra{-})}$ are the Pauli-$Z$ matrix and Pauli-$X$ matrix, respectively, where $\ket{-}=(\ket{0}-\ket{1})/\sqrt{2}$. The proposed networking teleportation protocol utilizes two different types of physical interaction of the cluster states, namely an $N$-qubit box-cluster state $\ket{C_{b, N}}$ (shown in Fig.~\ref{fig:fig_2}a, top) and a chain-type cluster state $\ket{C_{c, N}}$ (shown in Fig.~\ref{fig:fig_2}a, bottom). 

Notably, the advantages are twofold for the teleportation protocols to use either chain-type cluster states or box-cluster states. First, according to the connectivity map of 14-qubit \textit{ibmq\_16\_melbourne} device and the other quantum devices on IBMQ~\cite{ibmq}, chain-type cluster states and box-cluster states are the most feasible and applicable types of entangled states to be generated. Second, these two types of cluster states are natural resources to be integrated into different and complex networking protocols, such as universal measurement-based quantum computation~\cite{raussendorf2001one, briegel2009measurement, danos2006determinism, browne2007generalized}, error correction~\cite{gottesman1997stabilizer, schlingemann2001quantum, schlingemann2001stabilizer, aliferis2006simple}, blind quantum computation~\cite{broadbent2009universal, barz2012demonstration, morimae2013blind, greganti2016demonstration}, as well as quantum cryptography like quantum secret sharing~\cite{markham2008graph, bell2014experimental}.

\section{Networking teleportation protocols}
\begin{figure}[ht]
{\includegraphics[width=8.5cm]{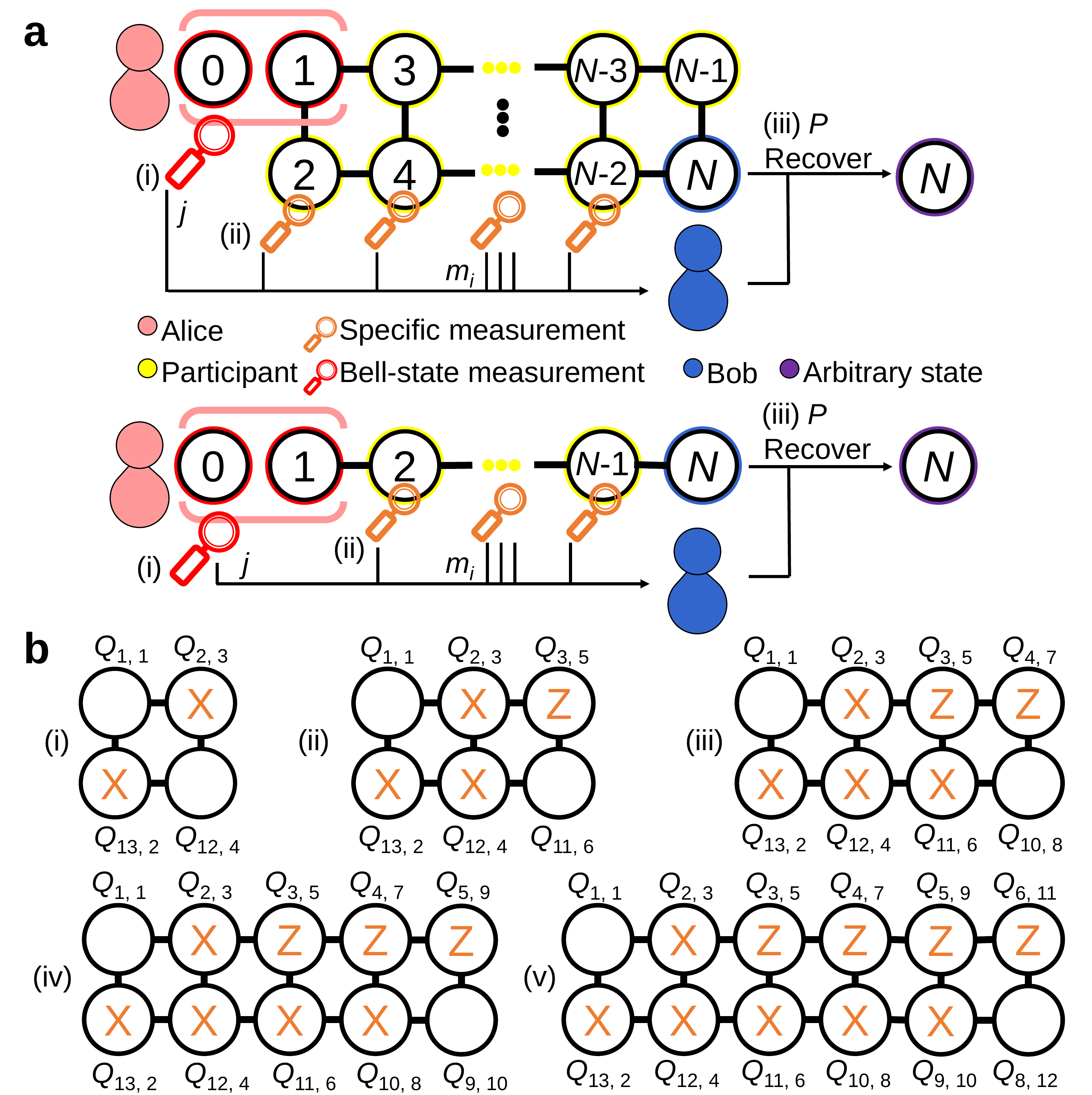}}%
 \caption{\label{fig:fig_2}%
Schematic of the $N$-qubit networking teleportation protocols and participants' specific measurement basis of $N$-qubit box-cluster state. (\textbf{a}) Schematic of the proposed networking teleportation protocols. The protocol aims to teleport an arbitrary state $\rho_{\rm{in}}$ (purple) using a shared $N$-qubit box-cluster state $\ket{C_{b, N}}$ (top) or $N$-qubit chain-type cluster state $\ket{C_{c, N}}$ (bottom), consisting of qubits 1--$N$ shared by Alice (pink), the participants (yellow), and Bob (blue). The protocol involves the following three steps: (\romannumeral1) Bell-state measurement performed on qubits 0 and 1 (red) at Alice's side, (\romannumeral2) local measurements on the specific basis performed on qubits 2--$N$-1 (orange) at the participants' side, and (\romannumeral3) unitary transformations $P$ (as shown in Eqs.~(\ref{eq:4}) and (\ref{eq:6})) on qubit $N$ (purple) to recover the teleported state at Bob's side. (\textbf{b}) Participants' specific measurement basis in the case of $\ket{C_{b, N}}$ with positive even integers from 4 to 12 (from diagrams (\romannumeral1) to (\romannumeral5), respectively). The notation $Q_{n,\,i}$ indicates that the logical qubit $i$ in the protocol description corresponds to the physical qubit $n$ on the 14-qubit \textit{ibmq\_16\_melbourne} device.
} 
\end{figure}

This section presents a general description of the proposed networking teleportation protocol, wherein either an $N$-qubit box-cluster state $\ket{C_{b, N}}$ (Fig.~\ref{fig:fig_2}a, top) with positive even integer $N$ up to 12 or an $N$-qubit chain-type cluster state $\ket{C_{c, N}}$ (Fig.~\ref{fig:fig_2}a, bottom) with arbitrary positive even integer $N$ (Fig.~\ref{fig:fig_2}a), is employed. The proposed protocols are applicable to arbitrary finite size of modules (Fig.~\ref{fig:fig_1}). It should be noted that Fig.~\ref{fig:fig_1} illustrates the concept of performing our protocols in a modular architecture, where the flying photon qubit is required for the Bell-state measurement performed by Alice because of the separation between the modules. However, it is worth stressing that our protocols are applicable regardless of whether modular architecture is used in universal superconducting quantum computer. Furthermore, the proposed protocols are also adaptable to other network communication systems, such as optics~\cite{wang2016experimental, barz2012demonstration, bell2014experimental}, ion traps~\cite{barreiro2011open, monz201114}, and NV centres~\cite{cramer2016repeated}.

In the following, we introduce the executive steps for Alice, the participants, and Bob, respectively in the proposed networking teleportation. 

\begin{figure*}
 {\includegraphics[width=17cm]{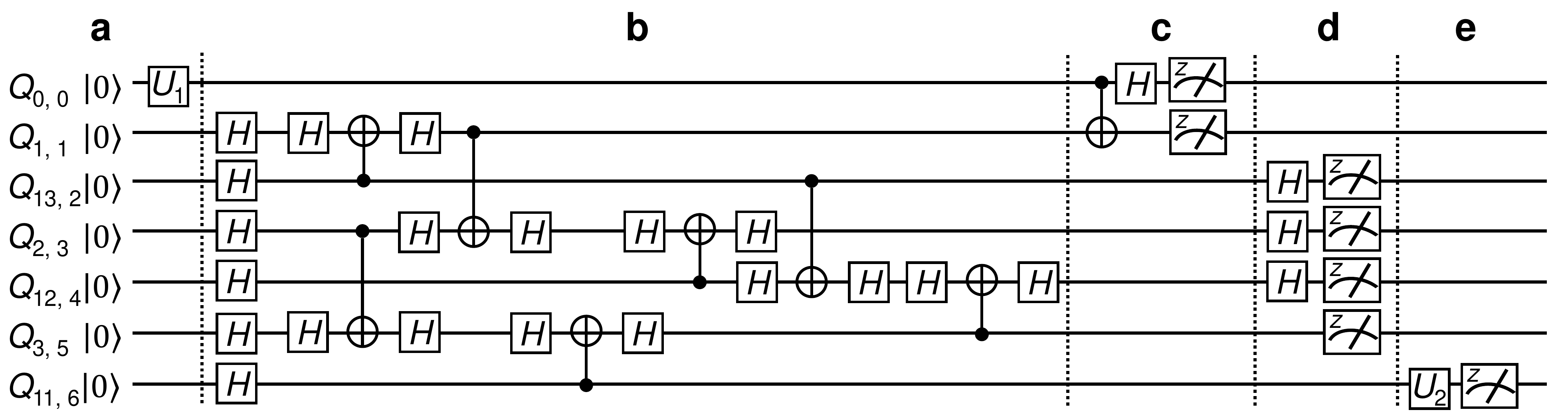}}%
 \caption{\label{fig:fig_3} 
Schematic showing quantum circuit of networking teleportation based on a 6-qubit box-cluster state $\ket{C_{b, 6}}$ on the \textit{ibmq\_16\_melbourne} device. The circuit to teleport an arbitrary state of $Q_{0,\,0}$ to $Q_{11,\,6}$ consists of the following: (\textbf{a}) all of the qubits are initialized to $\ket{0}$ and the arbitrary state of qubit $Q_{0,\,0}$ to be teleported is implemented by different unitary operations $U_1$; (\textbf{b}) $\ket{C_{b, 6}}$; (\textbf{c}) Bell-state measurement is performed on Alice's qubits $Q_{0,\,0}$ and $Q_{1,\,1}$; (\textbf{d}) local measurements are performed on the specific basis on the participants' qubits $Q_{13,\,2}$--$Q_{3,\,5}$; (\textbf{e}) state tomography is performed for the transported state on Bob's qubit $Q_{11,\,6}$, and is implemented by different unitary operations $U_2$ followed by measurement on the Pauli-$Z$ basis. (See the related text for a detailed description.)}
\end{figure*}

\begin{enumerate}     
     \item Measurement performed by Alice. Alice performs Bell-state measurement on qubit 0 (whose state $\rho_{\rm{in}}$ is arbitrary) and qubit 1, and obtains one of four possible results, namely 00, 01, 10, or 11. The measurement thus projects her two qubits into one of the four different Bell states $(U_j \otimes  I)\ket{\phi^+}$ with $j$=00, 01, 10, or 11, where 
\begin{eqnarray}
\begin{aligned}
\label{eq:3}
 &{U_{00}}=I, \hspace{1cm} &{U_{01}}=X,\\
 &{U_{10}}=Z, \hspace{1cm} &{U_{11}}=ZX,
\end{aligned}
\end{eqnarray}     
and $\ket{\phi^+} = (\ket{00}+\ket{11})/\sqrt{2}$. Alice communicates her measurement outcome $j$ to Bob (step (\romannumeral1) in Fig.~\ref{fig:fig_2}a). It is worth noting that Bell-state measurement (analysis) is one of the key element in quantum teleportation protocols and has been widely studied especially in optical systems~\cite{mattle1996dense, schuck2006complete, zhou2015complete, sheng2015two, zhou2016feasible, liu2019asymmetrical, zheng2019error, wang2019complete, li2019resource}; on the other hand, the Bell-state measurement performed on the universal IBM quantum computer is implemented by using universal logic gates followed by measurement on the Pauli-$Z$ basis (Fig.~\ref{fig:fig_3}c). 
    \item Local measurements performed by the participants. Each participant in the teleportation process performs measurement and communicates the result $m_{i}$ classically to Bob, where $m_{i} \in {\{+1,-1}\}$ represents the possible measurement outcome of the $i$-th participant's qubit on a specific measurement basis (step (\romannumeral2) in Fig.~\ref{fig:fig_2}a). The protocols based on $\ket{C_{b, N}}$ and $\ket{C_{c, N}}$ have $2^{(N-3)}$ and $2^{(N-2)}$ possible participant measurement results $(m_2,m_3,\ldots,m_i,\ldots,m_{N-1})$, respectively. The detailed steps of the participants' measurement processes for $\ket{C_{b, N}}$ and $\ket{C_{c, N}}$ are given in the following: 

    \begin{enumerate}
     \item For $\ket{C_{b, N}}$, every participant performs measurement on a specific basis on their qubit $i$. In particular, for all of the even qubits of the participants and qubit 3, measurement is performed on the Pauli-$X$ basis for measurements. By contrast, for the remaining odd qubits, measurement is performed on the Pauli-$Z$ basis for measurements (Fig.~\ref{fig:fig_2}b). A specific example is given in Methods section. 
     \item For $\ket{C_{c, N}}$, all of the participants perform local measurements on the Pauli-$X$ basis. 
     \end{enumerate} 
    \item Once Alice's and the participants' qubits have been collapsed by their measurements, Bob recovers $\rho_{\rm{in}}$ by applying appropriate unitary operations $P$ on his qubit $N$, where $P\in\{{I,Z,X,ZX,H,ZH,XH,ZXH}\}$. Note that $P$ is calculated based on Alice's measurement result $j$ and the participants' measurement results $(m_2,m_3,\ldots,m_i,\ldots,m_{N-1})$ (step (\romannumeral3) in Fig.~\ref{fig:fig_2}a). Bob’s operations for $\ket{C_{b, N}}$ and $\ket{C_{c, N}}$, respectively, are elaborated as follows:
    \begin{enumerate}
    \item For $\ket{C_{b, N}}$, Bob's operation to recover the arbitrary state has the form    
\begin{eqnarray}
\label{eq:4}
P=(\prod_{i=2}^{N-1}{O_{i,{m_{i}}}}){U_j}({H}^{(N)}), 
\end{eqnarray}     
where ${O_{i,{m_{i}}}}$ represents the unitary transformation according to measurement result $m_{i}$ of the $i$-th participant's qubit. The unitary operations $O_{i,{+1}}= {I}$ and $O_{i,{-1}}$ are defined as follows:
\begin{eqnarray}
\begin{aligned}
\label{eq:5}
 &{O_{2,{-1}}}=I, \hspace{1cm} &{O_{3,{-1}}}=X,\\
 &{O_{4,{-1}}}=Z, \hspace{1cm} &{O_{5,{-1}}}=I,\\
 &{O_{6,{-1}}}=X, \hspace{1cm} &{O_{7,{-1}}}=Z,\\
 &{O_{8,{-1}}}=Z, \hspace{1cm} &{O_{9,{-1}}}=X,\\
 &{O_{{10},{-1}}}=X, \hspace{1cm} &{O_{11,{-1}}}=Z.
\end{aligned}
\end{eqnarray}
Referring to Eq.~(\ref{eq:4}), ${U_j}$ is the unitary operation according to Alice's measurement result $j$, and is obtained using the definition given in Eq.~(\ref{eq:3}). In addition, ${H}^{(N)}$ is the Hadamard transformation ($H$) of qubit number $N$ and has a value of ${H}^{(N)}=I$ if qubit number $N$ can be divided by four; or ${H}^{(N)}=H$ otherwise. An illustrative case is given in Methods section. 

    \item For $\ket{C_{c, N}}$, Bob's operation to recover the arbitrary state is written in the form
\begin{eqnarray}
\label{eq:6}
P=(\prod_{i=2}^{N-1}{O_{{i},{m_{i}}}}){U_j}{H},
\end{eqnarray} 
where ${O_{{i},{m_{i}}}}$ is the unitary transformation according to measurement result $m_{i}$ of the $i$-th participant's qubit. The unitary operations ${O_{{i},{+1}}}=I$ and ${O_{{i},{-1}}}$ are defined respectively as follows:
\begin{eqnarray}
\begin{aligned}
\label{eq:7}
{O_{{i},{-1}}}=\left\{
\begin{array}{rcl}
X     &   & \rm{if}\ \it{i}\ \rm{is} \  \rm{even}\\
Z     &   & \rm{if}\ \it{i}\ \rm{is} \  \rm{odd}\\
\end{array} \right.. 
\end{aligned}
\end{eqnarray}

As described above, ${U_j}$ in Eq.~(\ref{eq:6}) is the unitary operation determined from Alice's measurement result $j$ using the definition given in Eq.~(\ref{eq:3}). An illustrative example is given in Methods.

    \end{enumerate}
\end{enumerate}        

It should be noted that when calculating $P$ using the protocols based on $\ket{C_{b, N}}$ or $\ket{C_{c, N}}$, there are four properties of matrix multiplication: (\romannumeral1) $IX=XI=X$, $IZ=ZI=Z$; (\romannumeral2) $XZX=Z$, $ZXZ=X$; (\romannumeral3) $XX=I$, $ZZ=I$, $II=I$; and (\romannumeral4) when $P$ is lastly calculated to be $XZ$, then $P$ should be considered to be $ZX$. Let us give four illustrative examples: (1) $P = IIZIXH \stackrel{(\rm{\romannumeral1})}{\longrightarrow} ZXH$; (2) $P = IZIIXIZH \stackrel{(\rm{\romannumeral1})}{\longrightarrow} ZXZH \stackrel{(\rm{\romannumeral2})}{\longrightarrow} XH$; (3) $P = IXZIXZZX \stackrel{(\rm{\romannumeral1})}{\longrightarrow} XZXZZX \stackrel{(\rm{\romannumeral2})}{\longrightarrow}  ZZZX \stackrel{(\rm{\romannumeral3})}{\longrightarrow} IZX \stackrel{(\rm{\romannumeral1})}{\longrightarrow}  ZX$; (4) $P = XZXZXZH \stackrel{(\rm{\romannumeral2})}{\longrightarrow}  ZZXZH \stackrel{(\rm{\romannumeral3})}{\longrightarrow}  IXZH \stackrel{(\rm{\romannumeral1})}{\longrightarrow}  XZH \stackrel{(\rm{\romannumeral4})}{\longrightarrow}  ZXH$.

It is worth noting that the two proposed protocols based on either a $\ket{C_{b, N}}$ or a $\ket{C_{c, N}}$ are both systematically scalable to arbitrary positive even integer $N$. To extend the proposed networking protocol based on $\ket{C_{b, N}}$ for $N>12$, the specific measurement basis of the $i$-th participant's qubit and the unitary operations $O_{i,{-1}}$ for $i>11$ are required to be defined. After all the measurements are performed by Alice and participants, Bob then calculates his operation $P$ using Eq.~(\ref{eq:4}) and recovers $\rho_{in}$ by applying $P$ to his qubit.

\section{Identifying quantum-mechanical process of networking teleportation}
\label{sec:C}
A generic classical process model \cite{hsieh2017quantifying} was utilized to quantitatively analyze the performance of a real quantum device, on which the proposed teleportation protocol was executed.

Suppose that a process of interest is created and its normalized process matrix, $\chi_{\rm{expt}}$, is obtained experimentally via the method of process tomography (PT) \cite{nielsen2002quantum}. If the experimental process cannot be described at all by any classical processes (denoted as $\chi_C$), then $\chi_{\rm{expt}}$ is said to be a genuinely quantum process (denoted by $\chi_Q$) \cite{hsieh2017quantifying}.  

A classical process, $\chi_C$, comprises a classical initial state and its evolution to a final state. The initial system can be regarded as a physical object with properties satisfying the assumption of realism. The system then evolves according to classical stochastic theory to a final state. It should be noted that the assumption of realism predicates that the system state can be described by a set of measurement outcomes. Moreover, the dynamics of these classical states are fully described by the transition probabilities from a specific state to a final state \cite{hsieh2017quantifying}.  

An experimental process, $\chi_{\rm{expt}}$, is identified as non-classical, i.e., close to the ideal quantum process $\chi_{Q_{I}}$, if the process fidelity satisfies that $F_{p}\equiv \text{tr}(\chi_{\rm{expt}}\chi_{Q_{I}})>F_{C}\equiv \mathop{max}\limits_{\chi_{C}} \text{tr}(\chi_{C}\chi_{Q_{I}})$ or ${\overline{F}}_{s,{\rm{expt}}}>{\overline{F}}_{s,C}$, where ${\overline{F}}_{s,{\rm{expt}}(C)}=(2F_{{\rm{expt}}(C)}+1)/3$ is the average state fidelity \cite{gilchrist2005distance}. It should be noted that ideal teleportation corresponds to an identity process, for which the process fidelity threshold is given by
\begin{eqnarray}
\label{eq:8}
{F}_{C}\sim 0.683,
\end{eqnarray}
and the average state fidelity threshold for teleportation is
\begin{eqnarray}
\label{eq:9}
{F}_{s,C}\sim 0.789.
\end{eqnarray}

In contrast, the measure-prepare strategy \cite{massar2005optimal} in which Alice directly measures her input state and Bob then prepares the output state accordingly is included as a special case in our generic classical-process model. It can be understood by the fact that the measure-prepare process is not an optimal classical teleportation: ${F}_{C} = 0.5 < 0.683$, and the average state fidelity is ${F}_{s,C} = 0.667 < 0.789$. Accordingly, our model provides the strictest criteria to evaluate whether an experimental teleportation can outperform classical mimicry. Our model is further proven in a quantitative way for evaluating an experimental teleportation process \cite{Chia}.

\section{Networking teleportation implemented on \textit{ibmq\_16\_melbourne} device}
\label{sec:D}
\label{sec:D}
The proposed networking teleportation protocol was experimentally implemented on the 14-qubit \textit{ibmq\_16\_melbourne} device (Fig.~\ref{fig:fig_2}b). Five different box-cluster states were considered, namely a 4-qubit box-cluster state $\ket{C_{b, 4}}$ consisting of qubits $Q_{1,\,1}$--$Q_{12,\,4}$ (Fig.~\ref{fig:fig_2}b(\romannumeral1)); a 6-qubit box-cluster state $\ket{C_{b, 6}}$ consisting of qubits $Q_{1,\,1}$--$Q_{11,\,6}$ (Fig.~\ref{fig:fig_2}b(\romannumeral2)); an 8-qubit box-cluster state $\ket{C_{b, 8}}$ consisting of qubits $Q_{1,\,1}$--$Q_{10,\,8}$ (Fig.~\ref{fig:fig_2}b(\romannumeral3)); a 10-qubit box-cluster state $\ket{C_{b, 10}}$ consisting of qubits $Q_{1,\,1}$--$Q_{9,\,10}$ (Fig.~\ref{fig:fig_2}b(\romannumeral4)); and a 12-qubit box-cluster state $\ket{C_{b, 12}}$ consisting of qubits $Q_{1,\,1}$--$Q_{8,\,12}$ (Fig.~\ref{fig:fig_2}b(\romannumeral5)).

Fig.~\ref{fig:fig_3} shows a schematic illustration of the implemented networking teleportation procedure for the 6-qubit box-cluster state $\ket{C_{b, 6}}$ (Fig.~\ref{fig:fig_1}c). As shown, all of the qubits $Q_{n,\,i}$ are initially prepared in the state $\ket{0}$ and the arbitrary quantum state of qubit $Q_{0,\,0}$ to be teleported is then prepared by applying the unitary operation $U_1$ (Fig.~\ref{fig:fig_3}a). To transport an arbitrary quantum state of qubit $Q_{0,\,0}$ to qubit $Q_{11,\,6}$, the networking teleportation procedure commences by preparing a 6-qubit cluster state using the definitions given in Eqs.~(\ref{eq:1}) and (\ref{eq:2}). The CZ gates are then implemented by a CNOT gate and two $H$ gates in accordance with the connectivity map of the 14-qubit \textit{ibmq\_16\_melbourne} quantum processor (Fig.~\ref{fig:fig_3}b).

Bell-state measurement is then performed on qubits $Q_{0,\,0}$ and $Q_{1,\,1}$ (Fig.~\ref{fig:fig_1}c(\romannumeral1)) using a CNOT gate and an $H$ gate; followed by measurement on the Pauli-$Z$ basis (Fig.~\ref{fig:fig_3}c). As described previously, the participants perform measurements on a specific basis (Fig.~\ref{fig:fig_1}c(\romannumeral2)). In particular, for all of the even qubits $Q_{13,\,2}$, $Q_{12,\,4}$ and the qubit $Q_{2,\,3}$, the participants perform measurements on the Pauli-$X$ basis (Fig.~\ref{fig:fig_2}b(\romannumeral2)), which is implemented by an $H$ gate, followed by measurement on the Pauli-$Z$ basis. For the remaining odd qubit, $Q_{3,\,5}$,  the participant performs measurement on the Pauli-$Z$ basis (Fig.~\ref{fig:fig_3}d). 

In the last step of the protocol, Alice sends her measurement result $j$, and each of the participants sends his or her measurement result $m_i$, to Bob through a classical communications channel. Bob then applies the last unitary operation $P$ defined in Eq.~(\ref{eq:6}) on qubit $Q_{11,\,6}$ to recover the transported state relying on the results from Alice and participants he received (Fig.~\ref{fig:fig_1}c(\romannumeral3)). 

Note that IBM Q Experience only permits at most one measurement on every given qubit. Moreover, no operations can be employed after a measurement. Thus, it is possible only to obtain the probabilities of all the possible measurement outcomes. 

To complete the networking teleportation procedure, quantum state tomography \cite{nielsen2002quantum} is performed on qubit $Q_{11,\,6}$ to reconstruct the density matrix of the teleported state, $\rho_{\rm{out}}$, by measuring the state in the Pauli basis \{$X$, $Y$, $Z$\}, where ${Y = (\ket{R}\bra{R}-\ket{L}\bra{L})}$ with $\ket{R}=(\ket{0}+i\ket{1})/\sqrt{2}$ and $\ket{L}=(\ket{0}-i\ket{1})/\sqrt{2}$ (Fig.~\ref{fig:fig_3}e). 

Experimentally, the measurements performed on the Pauli-$X$ or Pauli-$Y$ basis are implemented by using different transformations $U_2$ followed by measurement on the Pauli-$Z$ basis. In particular, the measurement on the Pauli-$X$ basis is implemented by an $H$ gate followed by measurement on the Pauli-$Z$ basis; while the measurement on the Pauli-$Y$ basis is implemented by $S^{\dag}$ and $H$ gates followed by measurement on the Pauli-$Z$ basis. Finally, the last unitary operation $P$ is applied as a post-selection to the experimental density matrices based on the measurement results informed by Alice and the participants, respectively. Herein, it is assumed that the transported state is perfectly recovered by Bob’s operation, $P$.  

To quantitatively characterize the performance of the real quantum processor, including all the required elements for teleportation, where the proposed protocol is implemented, the experimental results were evaluated using the process fidelity criterion (\ref{eq:8}). In performing the evaluation, complete PT~\cite{nielsen2002quantum} was applied to the teleported state $\rho_{\rm{out}}$ of the protocol. Furthermore, states $\rho_{\rm{in}} \in \{\ket{0}\bra{0},\ket{1}\bra{1},\ket{+}\bra{+},\ket{R}\bra{R}\}$ were chosen as the input states for teleportation. The teleportation process was described by the following positive Hermitian process matrix $\chi_{\rm{expt}}$:
\begin{eqnarray}
\label{eq:10}
\rho_{\rm{out}}=\sum_{m, n=1}^{4}\chi_{mn}M_m \rho_{\rm{in}} M_n,
\end{eqnarray}
where $M_1 = I$, $M_2 = X$, $M_3 = -iY$, and $M_4 = Z$. The ideal teleportation process matrix, $\chi_{Q_{I}}$, has only one non-zero element, $(\chi_{Q_{I}} )_{11}=1$. In other words, the input state is teleported without any loss in fidelity (Fig.~\ref{fig:fig_4}a).

Experimentally, to encode the qubit $Q_{0,\,0}$ to be teleported, where this qubit starts in the $\ket{0}$ state, various unitary gates $U_1$ are applied (Fig.~\ref{fig:fig_3}a). Specifically, to encode the $\ket{1}$ ($\ket{+}$) state, an $X$ ($H$) gate is placed on qubit $Q_{0,\,0}$, while to encode the $\ket{R}$ state, an $H$ gate is first applied followed by an $S$ gate, where $S=\ket{0}\bra{0}+i\ket{1}\bra{1}$.

\section{Identification of quantum-mechanical process of networking teleportation}
Before considering the identification of a real device, on which the proposed $N$-party networking teleportation protocol is conducted, the protocol was simulated on the 32-qubit \textit{ibmq\_qasm\_simulator} device, which enables anonymous users to compose ideal multi-shot executions of quantum circuits and then returns counts through IBM Q Experience \cite{ibmq}. The aim of the simulation was to verify the correctness of the individual steps in the proposed protocol based on an $N$-qubit box-cluster state $\ket{C_{b, N}}$ and a chain-type cluster state $\ket{C_{c, N}}$, respectively, for $N$ up to 12. 

\begin{figure}
 {\includegraphics[width=8.5cm]{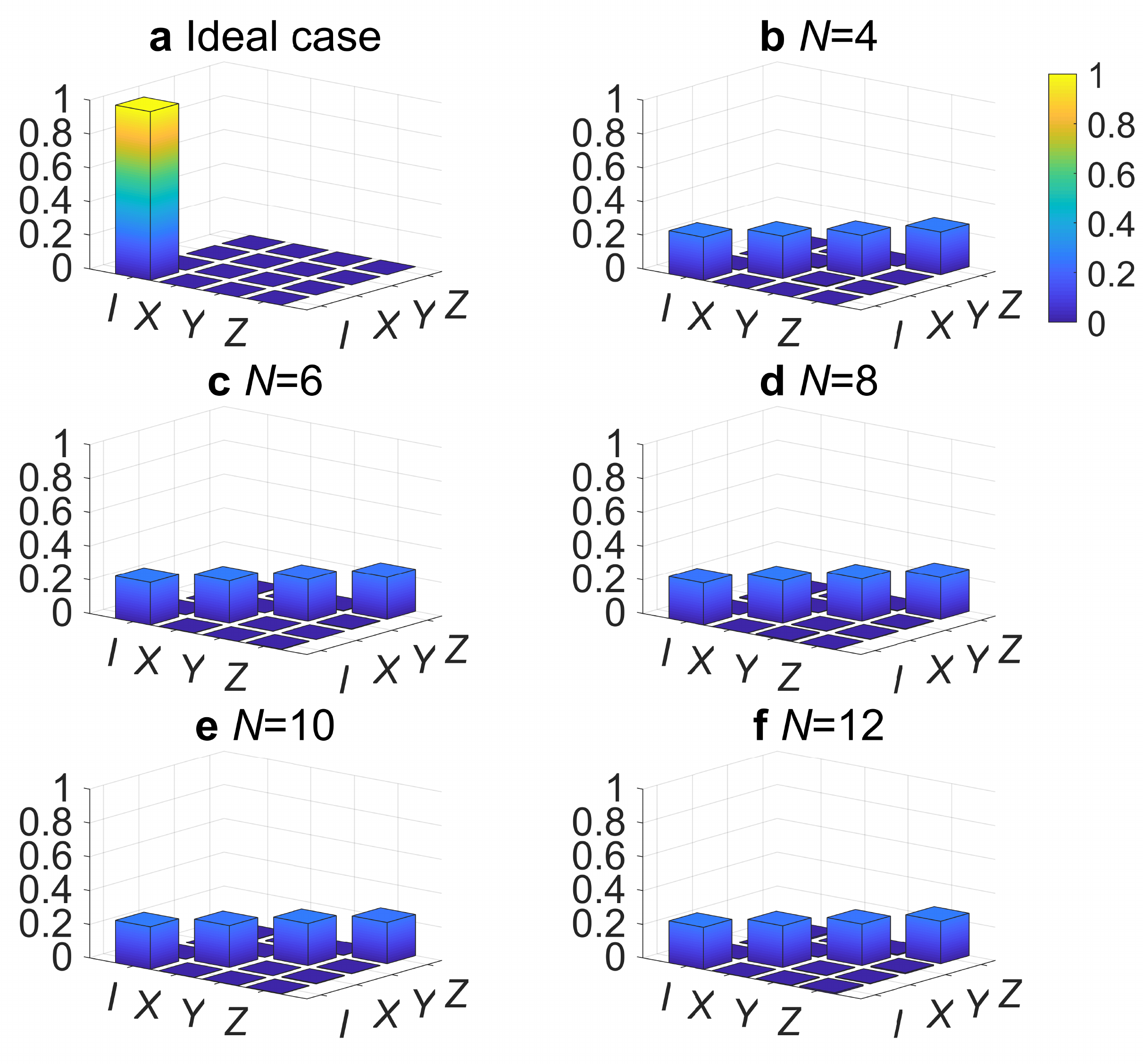}}
 \caption{\label{fig:fig_4} Absolute values of the reconstructed process matrix $\chi_{\it{mn}}$ with $m$, $n=$ $1$, $2$, $3$, $4$ for the proposed networking teleportation protocol based on $\ket{C_{b, N}}$ for: (\textbf{a}) ideal quantum teleportation; (\textbf{b}) $N$=4; (\textbf{c}) $N$=6; (\textbf{d}) $N$=8; (\textbf{e}) $N$=10; and (\textbf{f}) $N$=12.}
\end{figure}

\begin{figure}
 {\includegraphics[width=8.5cm]{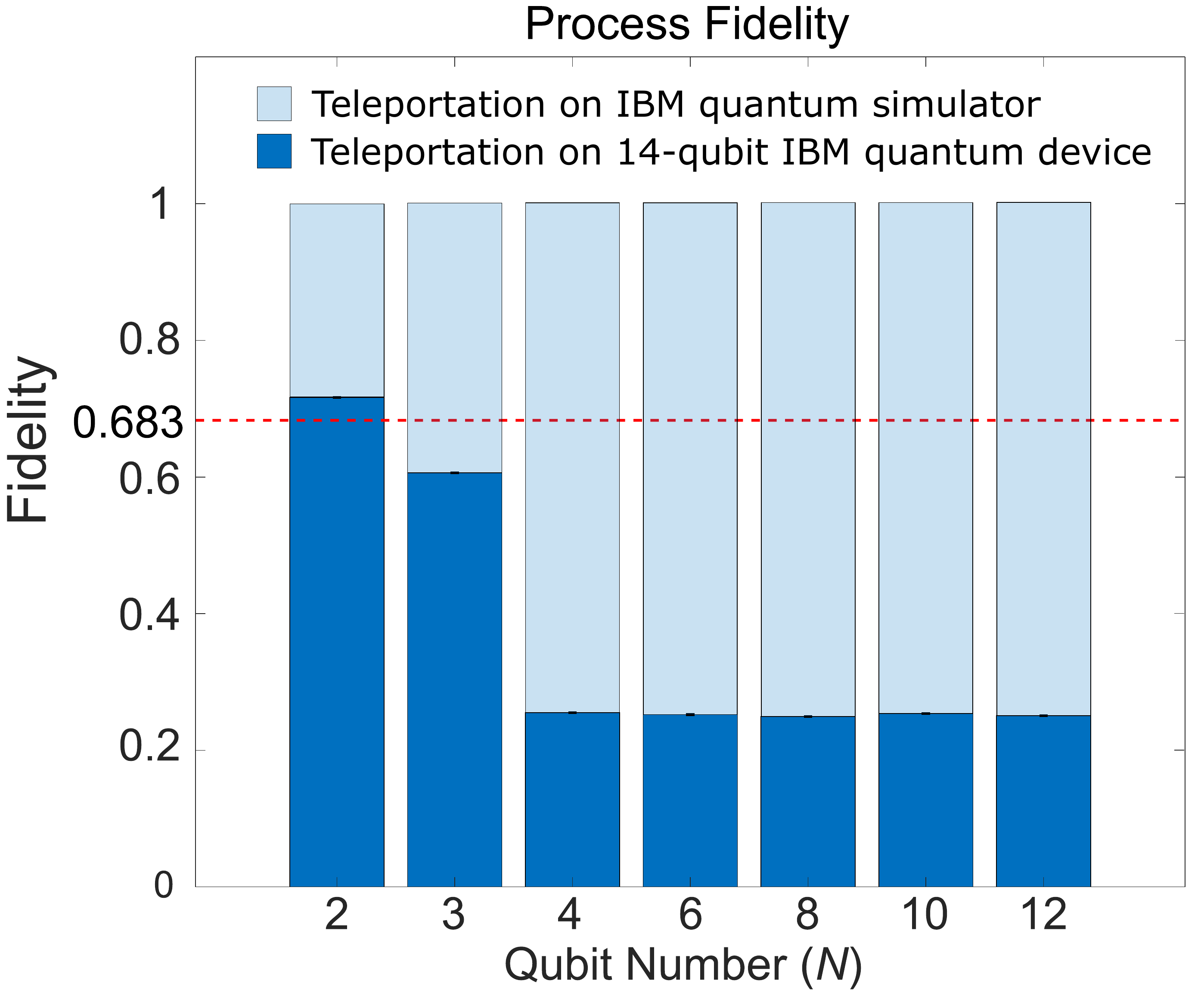}}
 \caption{\label{fig:fig_5} Process fidelity of $N$-qubit teleportation based on 2- and 3-qubit chain-type cluster states and $\ket{C_{b, N}}$ for qubit numbers $N$=4, 6, 8, 10 and 12, respectively. The red dotted line denotes the maximum classical process fidelity of $F_C=0.683$. The bars colored light blue and dark blue denote the results obtained from the 32-qubit \textit{ibmq\_qasm\_simulator} device and the 14-qubit \textit{ibmq\_16\_melbourne} device, respectively.}
\end{figure}

The process fidelities of the networking teleportation protocol based on $\ket{C_{b, N}}$ 32-qubit \textit{ibmq\_qasm\_simulator} device were calculated to be $F_{p}$ = $0.9972$, $1.0007$, $1.0032$, $0.9997$ and $0.9997$ for qubit numbers of $N$=4, 6, 8, 10 and 12, respectively (Fig.~\ref{fig:fig_5}). Meanwhile, the process fidelities of the networking teleportation protocol utilizing $\ket{C_{c, N}}$ were calculated to be $F_{p}$ = $0.9966$, $1.0004$, $0.9988$, $1.0004$ and $1.0004$, respectively. The state fidelities for the teleported quantum states $\rho_{\rm{out}}$ of the protocols based on $\ket{C_{b, N}}$ and $\ket{C_{c, N}}$, respectively, for $N$ up to 12 were calculated to be $F_{s}$ = $1.0000$ in both cases.

We can observe that the process fidelities of the protocols based on $\ket{C_{b, N}}$ for $N$=6 and $8$ and $\ket{C_{c, N}}$ for $N$=6, 10 and 12 are higher than 1, while those for the other values of $N$ are lower than 1. The reason is that if we conduct the simulation on the 32-qubit \textit{ibmq\_qasm\_simulator} device, we will obtain approximate probabilities. However, we will obtain an exact result if and only if the probabilities are zero or one. More specifically, events with zero probabilities will never be observed, while events with probability $0<p<1$ will be observed proportional to ``$p$'' (but unlikely to be exactly ``$p$'') \cite{Doi:2019:QCS:3310273.3323053}. The tomographic measurement of density matrices using this simulation measurement result may be able to produce results that violate important basic properties like positivity \cite{james2005measurement}. Therefore, the process fidelity is not exactly equal to 1 even though the state fidelities are all equal to 1.

Having validated the proposed protocol, it was conducted based on $\ket{C_{b, N}}$ for $N$=4 to 12 on the 14-qubit \textit{ibmq\_16\_melbourne} device. Complete PT was implemented on the teleported state $\rho_{\rm{out}}$, and formalisms and criteria described in Eqs.~(\ref{eq:11}) and (\ref{eq:12}) were used to evaluate the performance of the real device, on which our protocol was implemented. Figs.~\ref{fig:fig_4}b-f show the reconstructed process matrix $\chi_{\rm{expt}}$ for different $N$. One can observe that the experimental teleportation process matrix has four evenly distributed non-zero elements $(\chi_{\rm{expt}} )_{11}$, $(\chi_{\rm{expt}} )_{22}$, $(\chi_{\rm{expt}} )_{33}$, $(\chi_{\rm{expt}} )_{44}$, while ideally it should have only one non-zero element $(\chi_{Q_{I}} )_{11}=1$. In other words, the input state is teleported with nearly $75\%$ loss in fidelity.

To investigate the effect of the qubit number $N$ on the performance of the real processor, where the proposed protocol was conducted utilizing both a 2-qubit chain-type cluster state and a 3-qubit chain-type cluster state, respectively. The process fidelities $F_{p}\equiv \text{tr}(\chi_{\rm{expt}}\chi_{Q_{I}})$ were calculated to be $F_{p}$ = $0.7166\pm0.0010$,  $0.6063\pm0.0012$, $0.2550\pm0.0012$, $0.2523\pm0.0012$, $0.2493\pm0.0012$, $0.2539\pm0.0012$ and $0.2508\pm0.0012$ for qubit numbers $N$=2, 3, 4, 6, 8, 10 and 12, respectively (Fig.~\ref{fig:fig_5}). Note that each experimental value in Fig.~\ref{fig:fig_5} corresponds to the mean value obtained over 8192 measurements of 10 times. Note also that the error bars are obtained by Poissonian counting statistics and are rounded off to 4 decimals. Finally, the experimental values for $N$=4-12 and $N$=2-3 were accessed through IBM Q Experience \cite{ibmq} on 09 December 2018 and 25 August 2019, respectively.

From the experimental results obtained from the 14-qubit \textit{ibmq\_16\_melbourne} device reported above, one can observe that the experimental process fidelities for qubit numbers $N$=2, 3, 4, 6, 8, 10 and 12 decrease as $N$ increases. It is also observed that the quality of the experiments when utilizing a 3-qubit chain-type cluster state and $\ket{C_{b, N}}$ for qubit numbers $N$=4, 6, 8, 10 and 12 does not go beyond the maximum process fidelity of $F_{C} = 0.683$ (Eq.~\ref{eq:8}) that can be achieved classically (Fig.~\ref{fig:fig_5}). Finally, it is noted that the process fidelities are close to 0.25 for qubit numbers $N$=4, 6, 8, 10 and 12.

The state fidelity $F_s$ of the teleportation process is defined as the overlap of the ideal transported state $\rho_{\rm{in}}$ and the experimental density matrix $\rho_{\rm{out}}$. In other words, $F_s$($\rho_{\rm{in}}$, $\rho_{\rm{out}}$) = $\rm{tr}\sqrt{\sqrt{\rho_{\rm{in}}}\rho_{\rm{out}}\sqrt{\rho_{\rm{in}}}}$. It can be observed that the state fidelities of the four transported states  (shown in Table~\ref{tab:table1}) also do not surpass the maximum value of 0.789 in Eq.~\ref{eq:9} which is achievable by classical means. 

\begin{table}[ht]
\centering
\begin{tabular}{|l|l|l|}\hline
 $N$& $\ket{0}$          & $\ket{1}$ \\ \hline
 2  & $0.8692\pm0.0012$  & $0.8646\pm0.0012$ \\
 3  & $0.7168\pm0.0016$  & $0.7152\pm0.0016$ \\  
 4  & $0.5080\pm0.0017$  & $0.5055\pm0.0017$ \\  
 6  & $0.5034\pm0.0017$  & $0.4994\pm0.0017$\\ 
 8  & $0.4991\pm0.0017$  & $0.4957\pm0.0017$\\ 
 10 & $0.4984\pm0.0017$  & $0.5011\pm0.0017$\\ 
 12 & $0.5049\pm0.0017$  & $0.5031\pm0.0017$\\  \hline
 $N$ & $\ket{+}$         & $\ket{R}$ \\ \hline
 2  & $0.8252\pm0.0016$  & $0.7384\pm0.0017$ \\
 3  & $0.7453\pm0.0017$  & $0.7420\pm0.0017$ \\ 
 4  & $0.5000\pm0.0017$   & $0.5000\pm0.0017$ \\ 
 6  & $0.5000\pm0.0017$  & $0.5000\pm0.0017$ \\ 
 8  & $0.5000\pm0.0017$  & $0.5000\pm0.0017$ \\ 
 10 & $0.5000\pm0.0017$  & $0.5000\pm0.0017$ \\ 
 12 & $0.5000\pm0.0017$ & $0.5000\pm0.0017$ \\\hline
\end{tabular}
\caption{\label{tab:table1} State fidelities for the teleported quantum states $\rho_{\rm{out}}$ for qubit numbers $N$=2, 3, 4, 6, 8, 10 and 12, respectively. Each experimental value corresponds to the mean value obtained over 8192 measurements of 10 times, considers the Poissonian counting statistics, and is rounded off to 4 decimal places.}
\end{table}

In order to explore the potential causes of the experimental process fidelities of the real device, on which the proposed protocol was executed, the following section deconstructs the key ingredients required in the process. A series of analyses are additionally conducted on the shared entanglement and its fundamental CZ gate.

Firstly, we apply an optimal entanglement witness \cite{toth2005detecting} to detect the existence of genuine multipartite entangled states on the 14-qubit \textit{ibmq\_16\_melbourne} processor. The result implies that the existence of genuine multipartite entanglement cannot be detected in the experimental prepared state on the real quantum device (see Methods). Secondly, to clarify the effect of the CZ gate on the present experimental results for the multipartite cluster states, we characterized the CZ gate on the 14-qubit \textit{ibmq\_16\_melbourne} device by using PT. The experimental result suggests that an increasing number of $N$ and CZ gates leads to a corresponding decrease in the fidelity of the networking teleportation procedure on the 14-qubit quantum device (see Methods). Then we inquire into the effects of quantum noise in the experiments by comparing three common noise channels \cite{nielsen2002quantum}. The results shows that the noise in the networking teleportation process is similar to that produced in a depolarizing channel (see Methods).

\section{Discussion}
\noindent In this work, we have proposed two systematically scalable networking teleportation protocols consisting of $N$ parties utilizing either an $N$-qubit box-cluster state with positive even integer $N$ up to 12, or a chain-type cluster state with arbitrary positive even integer $N$, to transmit arbitrary quantum states inside and among the modules in a quantum network. The proposed protocols are adaptable to the benchmark provided by a generic classical-process model and applicable to arbitrary finite size of modules. Notably, the original teleportation protocol illustrates that two communication parties, Alice and Bob, can teleport the unknown state by sharing Einstein-Podolsky-Rosen (EPR) pairs~\cite{bennett1993teleporting}. Our protocols illustrate that many communication parties, Alice, participants and Bob, can teleport the unknown state by sharing multi-qubit cluster states. In contrast to the original protocol, the proposed protocols are more applicable for many communication parties in the future quantum network and can be further integrated into potential networking applications consisting of multiparties, such as protocols for quantum computation~\cite{raussendorf2001one, danos2006determinism, browne2007generalized, briegel2009measurement, gottesman1997stabilizer, schlingemann2001quantum, schlingemann2001stabilizer, aliferis2006simple, broadbent2009universal, barz2012demonstration, morimae2013blind, greganti2016demonstration} and quantum cryptography~\cite{markham2008graph, bell2014experimental}.

We have verified and tested the proposed protocols on both the IBM quantum simulator and the 14-qubit \textit{ibmq\_16\_melbourne} device. We have further utilized the generic classical-process model to quantify quantum-mechanical processes for identifying non-classical networking teleportation. The experimental results have shown that the process fidelities of the real quantum device, where the proposed networking teleportation protocol was conducted cannot go beyond the best mimicry attained by classical processes. That is, the components on the real device required for the networking teleportation process are not yet all qualified for use.

We have then unambiguously deconstructed the essential components in the networking teleportation process. We have prepared cluster states consisting of 4, 6, 8, 10 and 12 qubits on a 14-qubit \textit{ibmq\_16\_melbourne} device and have shown that genuine multipartite entanglement cannot be detected using entanglement witness operators. We then characterized the effect of the essential CZ gate on a 14-qubit \textit{ibmq\_16\_melbourne} device in constructing cluster states by process tomography and utilized the experimental process matrix of the controlled gate to reconstruct the whole networking teleportation procedure. The results showed that as the number of qubit $N$ and CZ gates increase, the fidelity of the networking teleportation procedure decreases. In addition, the noise in the experiments is close to that produced in a depolarizing channel. 

Qiskit is arranged in four libraries: Terra, Aqua, Aer and Ignis. The work presented herein utilizes Terra and Aer. Terra is intended for composing and optimizing quantum programs on a particular device, while Aer provides a simulator framework for users to compose and verify quantum circuits using the Qiskit software stack. In future studies, our work can further combine with error mitigation and correction software such as Ignis, one of the four libraries in Qiskit, to characterize the noise and errors induced by hardware via simulations.

Finally, through both the scalability to arbitrary finite even number of the qubit and the adaptability to the more general criteria for identifying non-classical teleportation of the proposed protocol, we provide an essential identification toolbox for future modular uses from a process point of view. It is worth stressing that the toolbox provides an essential assessment for identifying whether all the components on the real quantum device required in the networking teleportation process are all qualified for use. In particular, the proposed assessment method paves the way for further advancement of every key element in the whole networking teleportation process to facilitate the development of future modular techniques with improved reliability in performing quantum-information processing tasks.

\section{Methods}
\subsection{Illustrative examples for the steps in the proposed protocols.}
In the case of step 2(a), assume that the qubit number is $N=6$. In the measurement process, all of the even qubits (i.e., qubits 2 and 4) and qubit 3 are measured on the Pauli-$X$ basis, while the remaining odd particle (i.e., qubit 5) is measured on the Pauli-$Z$ basis (Fig.~\ref{fig:fig_1}c).

In the case of step 3(a), let us assume that qubit number $N$=6, Alice's measured outcome is $j=01$, and the measurement process for the participants' qubits yields $(m_2,m_3,m_4,m_5)=(-1,+1,-1,-1)$. According to the measurement results informed by Alice and the participants, $P$ = $IIZIXH$, In other words, $P=ZXH$ is applied to recover the input state $\rho_{\rm{in}}$. That is, Bob recovers $\rho_{\rm{in}}$ by applying first an $H$ gate, then an $X$ gate, and finally a $Z$ gate to his qubit.

In the case of step 3(b), we herein consider an illustrative example in which qubit number $N$=8, Alice's measured outcome is $j=10$, and the measurement results of the participants' qubits are $(m_2,m_3,m_4,m_5,m_6,m_7)=(+1,-1,+1,+1,-1,+1)$. According to the measurement results informed by Alice and the participants, $P=IZIIXIZH$. In other words, Bob applies $P=XH$ to recover the input state $\rho_{\rm{in}}$. That is, Bob recovers $\rho_{\rm{in}}$ by first applying an $H$ gate and then an $X$ gate.

\subsection{Detection of genuine multipartite entangled state.}
\label{sec:E1}
To detect the existence of genuine multi-partite entangled states on the 14-qubit \textit{ibmq\_16\_melbourne} device, which are the essential elements for realizing teleportation, we herein apply an optimal entanglement witness \cite{toth2005detecting} to evaluate the quality of the cluster states on the 14-qubit \textit{ibmq\_16\_melbourne} processor. For illustration purposes, we consider both a 6-qubit box-cluster state ${\ket{C_{b, 6}}}$ and a 6-qubit chain-type cluster state ${\ket{C_{c, 6}}}$. The witness for ${\ket{C_{b, 6}}}$ has the form
\begin{eqnarray}
\label{eq:11}
{\cal W}_{C_{b, 6}} = &&5I^{\otimes 6}-XZZIII-IZZXIZ-ZXIZII\nonumber\\
&&
-IIZIXZ-ZIXZZI-IIIZZX.
\end{eqnarray}
Meanwhile, the witness for ${\ket{C_{c, 6}}}$ has the form
\begin{eqnarray}
\label{eq:12}
{\cal W}_{C_{c, 6}} = &&5I^{\otimes 6}-XZIIII-IIZXZI-ZXZIII\nonumber\\
&&
-IIIZXZ-IZXZII-IIIIZX.
\end{eqnarray}

For a genuine 6-partite entanglement state close to ${\ket{C_{b, 6}}}$ (${\ket{C_{c, 6}}}$), $\left \langle{\cal W}\right \rangle$ is optimally equal to $-1$. To minimize the readout error caused by the measurements, the witnesses we used here only require two local measurement settings independent of the number of qubits for detection of each genuine multipartite entanglement. For example, $XZZXXZ$ and $ZXXZZX$ are required to evaluate ${\ket{C_{b, 6}}}$ , while $XZXZXZ$ and $ZXZXZX$ are required to evaluate ${\ket{C_{c, 6}}}$. 

Table~\ref{tab:table2} lists all the observables required to evaluate the witnesses for ${\ket{C_{b, 6}}}$ and ${\ket{C_{c, 6}}}$, respectively. Substituting the experimental results into Eqs.~(\ref{eq:11}) and (\ref{eq:12}) yields $\left \langle{\cal W}_{C_{b, 6}}\right \rangle$=5.126 and $\left \langle{\cal W}_{C_{c, 6}}\right \rangle$=4.1224. This result implies that the existence of genuine six-partite entanglement cannot be detected in the experimental prepared state on the real quantum device. In other words, it is necessary to improve the quality of multi-partite entanglement on the real quantum device. (Note that the experimental values shown in Table~\ref{tab:table2} were accessed through IBM Q Experience \cite{ibmq} on 20 December 2018.)

\begin{table}[ht]
\centering
\begin{tabular}{|l l|l l|}
\hline
$\ket{C_{b, 6}}$ & & $\ket{C_{c, 6}}$ &  \\ \hline
 Observable & Value & Observable & Value \\ \hline
 $XZZIII$ & $0.0050\pm0.0002$ & $XZIIII$ & $0.0811\pm0.0010$ \\
 $ZXZIII$ & $0.0020\pm0.0002$ & $ZXZIII$ & $-0.037$\\
 $ZIXZZI$ & $0.0018\pm0.0001$ & $IZXZII$ & $0.0139\pm0.0004$\\
 $IZZXIZ$ & $-0.087$          & $IIZXZI$ & $0.0193\pm0.0005$\\
 $IIZIXZ$ & $-0.03$           & $IIIZXZ$ & $0.3049\pm0.0016$\\
 $IIIZZX$ & $-0.017$          & $IIIIZX$ & $0.4951\pm0.0017$\\\hline
\end{tabular}
\caption{\label{tab:table2} Experimental values of all the observables on states $\ket{C_{b, 6}}$ and $\ket{C_{c, 6}}$, respectively, for entanglement witness ${\cal W}$ measurement. Each experimental value corresponds to the mean value obtained over 8192 measurements of 10 times and the error bars are obtained by Poissonian counting statistics.}
\end{table}

\subsection{Examination of experimental controlled gate.}
\label{sec:E2}
As shown in Eq.~\ref{eq:2}, the CZ gate is an essential entangling quantum gate for constructing a cluster state. To clarify the effect of the CZ gate on the present experimental results for the multipartite cluster states, the CZ gate on the 14-qubit \textit{ibmq\_16\_melbourne} device was fully characterized by means of quantum process tomography. In particular, the process matrix of the CZ gate was experimentally determined with maximum likelihood \cite{o2004quantum} and was then utilized to reconstruct the whole networking teleportation procedure utilizing $\ket{C_{b, 4}}$. The process fidelity was calculated to be $F_{p}$ = $0.2457$. In other words, this suggests that an increasing number of $N$ and CZ gates leads to a corresponding decrease in the fidelity of the networking teleportation procedure on the 14-qubit quantum device. (Note that the tomographic measurement of the CZ gate was accessed through IBM Q Experience \cite{ibmq} on 10 April 2019.)

To inquire into the effects of quantum noise in the experiments, we compared three common noise channels \cite{nielsen2002quantum}, namely a depolarizing channel $(\chi_{\text{D}})$, a phase damping channel $(\chi_{\text{AD}})$ , and an amplitude damping channel $(\chi_{\text{PD}})$, to the experimental process matrix $\chi_{\rm{expt}}$. The noise channels were defined respectively as

\begin{eqnarray}
\label{eq:13}
\chi_{\text{D}}(\rho)&=&(1-\frac{3}{4})I\rho I+\frac{1}{4}(X\rho X+Y\rho Y+Z\rho Z), \nonumber\\
\chi_{\text{PD}}(\rho)&=&(1-\frac{1}{2})I\rho I+\frac{1}{2}Z\rho Z,\\
\chi_{\text{AD}}(\rho)&=&(\frac{1}{2}I+\frac{1}{2}Z)\rho(\frac{1}{2}I\!+\frac{1}{2}Z)+\frac{1}{4}(X\!+\!iY)\rho (X\!-\!iY). \nonumber
\end{eqnarray}

An inspection of the computed fidelity values $F$($\chi_{\rm{expt}}$, $\chi_{\text{noise}})$ = $\rm{tr}\sqrt{\sqrt{\chi_{\text{noise}}}\chi_{\rm{expt}}\sqrt{\chi_{\text{noise}}}}$ (Table~\ref{tab:table3}) shows that the noise in the networking teleportation process is similar to that produced in a depolarizing channel. This then explains why the experimental process fidelities are all close to $0.25$.

\begin{table}[ht]
\centering
\begin{tabular}{|l|l|l|l|}
\hline
\textrm{$N$}&
\textrm{$\chi_{\text{D}}$}&
\textrm{$\chi_{\text{AD}}$}&
\textrm{$\chi_{\text{PD}}$}\\ \hline
4  & $0.9999$ & $0.7080$ & $0.7119$\\
6  & $1.0000$ & $0.7085$ & $0.7081$\\
8  & $0.9999$ & $0.7083$ & $0.7053$\\
10 & $1.0000$ & $0.7062$ & $0.7069$\\
12 & $0.9999$ & $0.7077$ & $0.7098$\\ \hline
\end{tabular}
\caption{\label{tab:table3}%
$F$($\chi_{\rm{expt}}$, $\chi_{\text{noise}}$) for networking teleportation protocol utilizing an $N$-qubit box-cluster state $\ket{C_{b, N}}$ and three common noise channels, respectively.}
\end{table}

\begin{acknowledgments}
The authors acknowledge the following: the IBM Q team for the access to their 14-qubit quantum computer through a cloud-computing interface. We also gratefully acknowledge the IBM Q team members, Rudy Raymond, Takashi Imamichi, and Chun-Fu Chen, for their valuable discussions and correspondence on implementation of the networking experiments through the cloud. Finally, we acknowledge the fruitful discussions with Wei-Ting Lee, Shih-Hsuan Chen, Chia-Kuo Chen, and Chien-Ying Huang. This work was partially supported by the Ministry of Science and Technology, Taiwan, under Grant Numbers MOST 107-2628-M-006-001-MY4 and MOST 107-2627-E-006-001. 

\end{acknowledgments}

\end{document}